\documentclass[aps,pra]{revtex4-2}
\usepackage{amsmath,amssymb,amsthm}
\usepackage[colorlinks=true,citecolor=blue,urlcolor=blue]{hyperref}
\usepackage{times,txfonts}
\usepackage{braket}
\usepackage{color}
\usepackage{natbib}
\usepackage{amsmath,blkarray}
\usepackage{mathtools}
\usepackage{latexsym}
\usepackage{tabularx, booktabs}
\usepackage{float}
\usepackage{amsfonts}
\usepackage{subcaption}
\usepackage{color,soul}
\begin{document}
\title{L$\ddot{u}$ders bounds of Leggett-Garg inequalities, $\mathcal{PT}$- symmetric evolution and arrow-of-time}
\author{Asmita Kumari}
\affiliation{Harish-Chandra Research Institute, Allahabad 211091, India}
\author{A. K. Pan}
\email{akp@nitp.ac.in}

\affiliation{National Institute Technology Patna, Ashok Rajhpath, Patna, Bihar 800005, India}

\begin{abstract}
 Leggett Garg inequalities (LGIs) test the incompatibility between the notion of macrorealism and quantum mechanics. For unitary dynamics, the optimal quantum violation of an LGI is constrained by the L$\ddot{u}$ders bound. However, the LGIs does not provide the necessary and sufficient for macrorealism. A suitably formulated set of no-signaling in time (NSIT) conditions along with the arrow-of-time (AOT) condition provides the same.  In this paper, we study two formulations in the three-time LG scenario, viz., the standard LGIs and the recently formulated variant of LGIs when the system evolves under $\mathcal{PT}$-symmetric Hamiltonian. We first demonstrate that the quantum violations of both forms of LGIs exceed their respective L$\ddot{u}$ders bounds and can even reach their algebraic maximum.  We further show that for the case of standard LGI, the violation of L$\ddot{u}$ders bound can be obtained when both NSIT and AOT conditions are violated. Interestingly, for the case of a variant of LGI, for suitable choices of relevant parameters, the quantum violation can even be obtained when only the AOT is violated but all NSIT conditions are satisfied. This feature has not hitherto been demonstrated. We discuss the further implication of our study.
\end{abstract}

\maketitle

\section{Introduction}
Since the inception of quantum mechanics (QM), it remains an open question how our everyday world realist view of the macroscopic world emerges from quantum formalism.  Schr$\ddot{o}$dinger \cite{sch} was the first to put forward this question through his famous cat experiment. Since then, this issue has been studied from various perspectives \cite{zur, bruk,ghi}. Motivated by Bell's theorem \cite{bell64}, in 1985, Leggett and Garg \cite{lg85,lg02} provided a refined definition of macrorealism and formulated a class of inequalities (henceforth, LGIs)  valid for any macrorealistic theory, thereby providing an elegant route for examining the status of macrorealism in QM.  The notion of macrorealism advocated by Leggett and Garg comprises the following three assumptions.
	
\emph{macrorealism per se:} If a macroscopic system has two or more macroscopically distinguishable ontic states available to it, the system remains in one of those states at all instant of time.
	
\emph{Non-invasive measurement:} The definite ontic state of the macrosystem is determined without affecting the state itself or its possible subsequent dynamics.

\textit{Induction :} It says that the outcome of a measurement on a system cannot be affected by its future measurement i,e., retro-causality is not allowed.

This assumption may also be termed as the arrow-of-time (AOT) condition and is commonly assumed to be satisfied by a physical theory. The violation of it leads retro-causality \cite{leifer,now,spe}. AOT condition plays an important role in our present study.

 The notion of LGIs have been extensively studied both theoretically \cite{kofler08,emary12,wild,kofler13,maroney14,emary,budroni14,clemente15,  budroni15, saha15,halli16,halli17,swati17,pan17,akumari18,pan18,halli19,halli19a,halli20,pan20,halli21} and experimentally \cite{lambert,goggin11,knee12,laloy10,george13,knee16,kati1,wang02}. There is a common perception that the LGI is \emph{temporal analog} of Bell's inequalities. This inference is motivated by the structural resemblance between Bell-CHSH inequalities and four-time LGIs. However, LG test shows serious pathology \cite{clemente15,clemente16,halli16,swati17,halli17,pan17,akumari18,pan18,halli19,halli19a,wild} in comparison to a Bell test. The no-signaling in space condition is always satisfied in any physical theory. Thus, the violation of a given Bell's inequality proves a failure of local realism and not the locality alone. In contrast, in the statistical version of the non-invasive measurability assumption, the no-signaling in time (NSIT) can be shown to be violated in QM. Thus, a macrorealist may argue that quantum violation arises due to violation of the NSIT condition. Hence, to draw meaningful conclusions from a LG test, one must satisfy the NSIT condition in QM by adopting a suitable measurement scheme.  Clemente with Kofler \cite{clemente15} proposed that suitable conjunction of  NSIT and AOT conditions is necessary for three-time LGIs. However, AOT is trivially satisfied in unitary QM. 

In this work, we restrict ourselves to a three-time LG scenario. We consider two different LGIs for our study, viz, standard LGIs and variant of LGIs, the latter is recently formulated by one of us \cite{pan18}. The optimal quantum value of the standard LGI in unitary dynamics is restricted to $1.5$, which we call L$\ddot{u}$ders bound. The L$\ddot{u}$ders bound for the variant of LGI in the three-time scenario is $2$. However, this can approach the algebraic maximum of the LG expression if $n$-time measurement scenario is taken into account  \cite{pan18}. We note here that, in \cite{budroni14} it was shown that the quantum violation of three-time LGIs could reach the algebraic maximum of the LG expression for a dichotomic observable of the spin-$j$ system if instead of L$\ddot{u}$ders rule the degeneracy breaking von Neumann rule is invoked. However, the relevance of that result in the context of LG scenario has been criticized in \cite{akumari18}.

Here, instead of unitary evolution, we consider the case when the system is evolved under $PT$-symmetric Hamiltonian \cite{bender02}. This provides an alternative way of looking at the violation of LGIs for the non-Hermitian system. Although the experimental realization of $PT$-symmetric QM is challenging, single-mode oscillation based on  $PT$-symmetric operation of an optoelectronic oscillator is achieved \cite{zhang18}. Very recently, Wu \emph{et al} \cite{wu19} has developed an elegant technique to dilate  $PT$-symmetric  Hamiltonian into Hermitian and tested experimentally by using a nitrogen-vacancy center system. In recent years, the $PT$-symmetric QM has gained considerably high attention for demonstrating various forms of strange quantumness, such as, increment of entanglement and violation of no-signalling principle \cite{b0,b01,b1,b2,b3,b4,b5,b6,b7,b8,b9,b10,pati19,pati14}, and many more. The reason for showing such a strange phenomenon can be attributed to the improper use of one of the following two assumptions \cite{b01} on which  $PT$-symmetric QM is built. First, the local quantum system is described by a $PT$-symmetric Hamiltonian, and it can coexist with a conventional quantum system. Second, post-measurement probability distributions are computed using conventionally normalized quantum states.

Recently, using $PT$-symmetric evolution \cite{bender02,asmita20}, it has been shown that the quantum violation of standard LGIs in QM can approach the algebraic maximum for four-time \cite{Karthik} as well as three-time LGI \cite{javid20} scenarios.  Although our work is along this line, we additionally study the quantum violation of the variant of LGIs in a three-time LG scenario. We demonstrate that to obtain the violation of LGI, at least one of the NSIT conditions should be violated.  To explicitly demonstrate this feature, we cast the LGIs in terms of the degree of NSIT and AOT conditions violations. This enables us to show that mere violation of one or more NSIT conditions is not enough to warrant the violation of LGIs, but one of the AOT conditions needs to be violated. In other words, violation of variant of LGI when the system evolved under $PT$-symmetric Hamiltonian the future measurement can affect the past measurements. 

The paper is organized as follows.  We introduce preliminaries corresponding to standard and variant of LGIs and $\mathcal{PT}$-symmetric Hamiltonian in Sec. II. In order to make our paper self-contained, we have also recapitulated the known result that the local operation of \(PT\)-symmetric Hamiltonians on bipartite entangled state violates the no-signaling principle.  In Sec. III, we have discussed the possibility of violating the AOT condition when the system evolves under $\mathcal{PT}$-symmetric evolution. Sec. IV demonstrated that quantum violations of both forms of LGIs reach their algebraic maximum when the system evolves under $\mathcal{PT}$-symmetric evolution. In Sec. V, we critically examine the implications for the quantum violations of L$\ddot{u}$ders bound of standard and variant of LGIs using $\mathcal{PT}$-symmetric evolution. Finally, in Sec. VI, we have provided a summary and discussion of our results.

\section{Preliminaries}
Let us briefly encapsulate the notion of LGIs and $PT$-symmetric evolution.
\subsection{Quantum violations of various LGIs for unitary dynamics}
Let us consider a dichotomic observable $M_i$ having outcomes $m_i = \pm 1$ at any instant of time $t_i$. Then the correlation function between the observables $M_i$ and $M_j$ measured at time $t_i$ and $t_j$ respectively is defined as
\begin{eqnarray}
\label{mij}
 \langle M_i M_j \rangle = \sum_{m_{i},m_{j} = \pm 1} m_{i} m_{j} P(m_{i}, m_{j})
\end{eqnarray}
In three-time LG scenario, an observable $M_1$ is measured at three different time $t_1, t_2$ and $t_3$ ($t_3 > t_2 > t_1$) which corresponds to the observables $M_{1}, M_{2}$ and $M_{3}$ respectively in unitary dynamics case.  One of the standard LGIs can be written as 
\begin{eqnarray}
\label{lgi1}
L_{13}&=& \langle M_{1} M_{2}\rangle + \langle M_{2} M_{3}\rangle - \langle M_{1} M_{3}\rangle \leq 1
\end{eqnarray}
which is assumed to be valid for a macrorealistic theory satisfying MRps, NIM and AOT. By relabeling the measurement outcomes of each $M_i$ as $M_i=-M_i$ with $i=1,2$ and $3$, three more standard LGIs can be obtained.

Standard LGIs are a particular class of inequalities obeying the assumption of MRps and NIM. It has been argued in \cite{pan18} that due to the sequential nature of the measurement in LGIs, LG scenario is more flexible than the Bell scenario. Hence, by keeping the assumptions of MRps and NIM intact, there remains a possibility to propose more variants of LGIs. Instead of two-time correlation functions in two-time LGIs scenario, in a variant of LGIs consists of three-time  correlation function $\langle M_{1} M_{2} M_3\rangle$, two-time correlation function $\langle  M_{i} M_j \rangle$ and $\langle  M_{k}\rangle $ is given by  
\begin{eqnarray}
\label{vlgi1}
V_k=-\langle M_{1} M_{2} M_3\rangle + \langle M_{i} M_{j}\rangle +\langle M_{k}\rangle \leq 1
\end{eqnarray}  
where $i,j,k=1, 2, 3$ with $j > i$.
Again by relabeling the measurement outcomes of each $M_{i}$ as $M_{i}=-M_{i}$ with $i=1,2$ and $3$, more variant of LGIs can be obtained.

In order to demonstrate the quantum violation, let us assume that the initial state of the system and  observable to be measured at time $t_1$ is 
\begin{equation}
\label{state}
|\psi(t_1)\rangle = \cos\theta |0 \rangle + e^{i\phi} \sin\theta |1 \rangle
\end{equation}
 and $M_1=\hat{\sigma_z}$ respectively. Then, the observables at time $t_2$ and $t_3$ are $M_{2}=U_{\Delta{|t_2-t_1|}} M_{1} U^{\dagger}_{\Delta{|t_2-t_1|}}$ and $M_{3}=U_{\Delta{|t_3-t_2|}} M_{2} U^{\dagger}_{\Delta{|t_3-t_2|}}$ respectively. Here  $U_{\Delta{|t_2-t_1|}}=U_{\Delta{|t_3-t_2|}} = \exp ( i \tau \hat{\sigma_x})$ such that $ t = g(t_2-t_1) =g(t_3-t_2)$. Then the quantum expression of LGI,  $L_{13}$ in Eq.(\ref{lgi1}) can be written as
\begin{eqnarray}
\label{qlgi}
(L_{13})_{Q}=2 \cos(2 t) -\cos(4 t)
\end{eqnarray}
which is state-independent. The optimal quantum value of $(L_{13})_Q$ is $1.5$ (L$\ddot{u}$ders bound) obtained at $ t=\pi/6$. In \cite{budroni13} it has been shown that such L$\ddot{u}$ders bound is irrespective of system size. 
Next, for the same state, as given in Eq.(\ref{state}) dichotomic observable $\sigma_z$  and the unitary evolution as used for standard LGI, the quantum expression of $V_{3}$ is obtained as
\begin{eqnarray}
\nonumber
(V_{3})_Q = \cos(2t)(1 + 4 \sin^2(t) \cos(2 \theta)) + 2  \sin^2(t) \cos(2 \theta) - \sin(4 t) \sin(2 \theta)\sin(\phi)
\end{eqnarray}
which is state dependent unlike the case of standard LGI. At $t=0.41, \theta = 2.66$ and $\phi = \pi/2$, the optimal quantum value of $(V_{3})_Q$ is $1.93$. It can even approach $2$ if one takes different value of coupling constants for evolutions between measurements.

\subsection{$\mathcal{PT}$-symmetric evolution}
An operator is said to be Hermitian if the eigenvalues of that operator are real and the corresponding eigenstates are orthogonal, and the evolution generated by it is unitary. The unitary evolution in QM is responsible for the conservation of probability. It should be noted that real eigenvalue is not only restricted to the Hermitian operator. There exists non-Hermitian operators having real eigenvalues, known as $\mathcal{PT}$-symmetric having symmetry property of time and parity. The well known $\mathcal{PT}$-symmetric non-Hermitian operator with respect to space-time reflection symmetry \cite{bender98,bender02,brody12} is given by 
\begin{eqnarray} 
\label{pt}
H = s \left( \begin{array}{cc} i \sin(\alpha) & 1 \\ 1 & -i \sin(\alpha) \end{array} \right)
\end{eqnarray}
where $s$ is scaling constant and $\alpha$ degree of Hermiticity of the Hamiltonian $H$. The Hamiltonian $H$ is non-Hermitian when $\alpha \neq 0$ and reduces to Hermitian one when $\alpha = 0$. In the present work, we are only concerned in the range of $H$ having  real eigenvalues, which can be obtained when $|\alpha| < \pi/2$. The real eigenvalues of $H$ are $E_{\pm} = \pm s \cos{\alpha}$ corresponding to the eigen-state 
\begin{eqnarray} 
|E_{\pm} (\alpha) \rangle =\frac{e^{ \pm i \alpha / 2}}{\sqrt{2 \cos{\alpha}}}  \left( \begin{array}{cc} 1 \\  \pm e^{\mp i \alpha} \end{array} \right)
\end{eqnarray}
respectively. It should be noted that unlike the eigenstates of Hermitian operator, $|E_{+} (\alpha) \rangle$ and $|E_{-} (\alpha) \rangle$ are non-orthogonal state. The specific point of $\alpha = \pm \pi /2$ at which eigenvalues becomes equal and  eigenstates $|E_{+} (\alpha) \rangle$ and $|E_{-} (\alpha) \rangle$  coalesce is known as exceptional point \cite{b01}. 

The time evolution that is governed by $\mathcal{PT}$-symmetric Hamiltonian $H$ is non-unitary and  is given by 
\begin{eqnarray}
\label{evl} 
U_{\alpha}(t) =e^{-i H \tau} = \frac{1}{\cos(\alpha)} \left( \begin{array}{cc}  \cos(t - \alpha) & -i \sin(t)  \\ i \sin(t) & \cos(t + \alpha) \end{array} \right)
\end{eqnarray}
where $t= s \tau \cos(\alpha)$ and we assume $\hbar = 1$. 
We use the $\mathcal{PT}$-symmetric evolution between two measurements.

\subsection{Bipartite entangled state and the argument of the violation of no-signaling principle}
\label{tin}

In \cite{b01} it has been shown that local operation of \(PT\)-symmetric Hamiltonians on bipartite entangled state violates the no-signaling principle. In order to make our paper self-contained, in this section, we recapitulate the known results of violation of the no-signaling principle.

 Let Alice and Bob are two parties situated in two space-like separated locations and share a maximally entangled state 
\begin{eqnarray}
|\psi \rangle_{AB} = \frac{1}{\sqrt{2}}(|00 \rangle + |11 \rangle),
\end{eqnarray}
where \(|0\rangle\) and \(|1\rangle\) are eigenstates of the Pauli  operator $\sigma_z$.
Under any unitary dynamics by Alice, the reduced density matrix of Bob is $ \rho_{B} = \mathbb{I}/2 $.

This is a form of the no-signaling condition within QM, which says that the probability distribution of outcomes of Bob is unaffected by Alice's choice of quantum operations. Indeed, the no-signaling naturally holds 
 if Alice chooses to perform a unitary operation on an augmented system on her side. 

 If the system belongs to Alice's sub-system evolves under $U_{\alpha}$ as given in Eq.(\ref{evl}),
 then the composite state becomes 
\begin{eqnarray}
|\psi \rangle^{\alpha}_{AB} = U_{\alpha}(t) \otimes \mathbb{I} \left(\frac{1}{\sqrt{2}}\left(|00 \rangle + |11 \rangle\right)\right) 
\end{eqnarray}
where \(\mathbb{I}\) denotes the identity operator acting on Bob's sub-system. The composite density matrix then becomes

$\rho_{AB}^{\alpha} = U_{\alpha}(t) \otimes \mathbb{I} \rho_{AB} U_{\alpha}(t)^{\dagger} \otimes  \mathbb{I}$.

 Now, by taking the partial trace over Alice's sub-system, the reduced state of Bob ( after renormalization), is obtained as
\begin{eqnarray}
\rho_{B}^{\alpha} &=& \mbox{Tr}_A[\rho_{AB}^{\alpha}] = \frac{1}{N_1} \left(
\begin{array}{cc}
  b_1 & b_4 \\
 b_3 & b_2 \\
\end{array}
\right), 
\end{eqnarray}
where $N_1 = 2 \sec ^2(\alpha)  \sin ^2( t)+\cos (2 t)$, and
\begin{eqnarray}
\nonumber
b_1 &=&  
\frac{1}{2} \sec ^2(\alpha ) (\cos (2 (\alpha -t))-\cos (2 t)+2), \ \ b_2 = 
\frac{1}{2} \sec ^2(\alpha ) (\cos (2 (\alpha +t))-\cos (2 t)+2), \\ \nonumber
b_3 &=& 
-2 i \tan (\alpha ) \sec (\alpha ) \sin ^2(t), \ \  
b_4 =
2 i \tan (\alpha ) \sec (\alpha ) \sin ^2(t).
\end{eqnarray}
The no-signaling principle remain intact if  $\rho^{\alpha}_{B} = \mathbb{I}/2$. However, this happens only if \(\alpha =0\) or \(t =0\). Note that, \(\alpha =0\) leads to a hermitian \(H\) and \(t=0\)  implies that the evolution did not happen. Hence,  local operation of  \(PT\)-symmetric Hamiltonians can be used to violate the no-signaling principle. At \(\alpha = \pm \pi/2\), the eigenvectors of the \(PT\)-symmetric Hamiltonian coalesce and eigenstates become same. So, except for the case when \(\alpha = \pm \pi/2\), all other \(PT\)-symmetric non-hermitian Hamiltonians with real eigenvalues lead to signaling  \cite{b01}.  Since, the signaling in the physical thing is not possible the solution of the above problem comes as following.  In this paper we are restricted ourself for $|\alpha| < \pi/2$.

\section{$PT$ symmetric evolution and Arrow-of-time}
We show here that for sequential measurement scenario, when the system evolves under $U_{\alpha}$ between two measurements, there is a possibility of retro-causality, implying that the future measurements can influence the prior measurements.

Let initial density matrix of the system at time $t_j$ is $\rho_j$.  We perform the measurement of an observable $M_j$ at $t_j$ and $t_k$ with $t_j<t_k$. The system evolves under \(PT\)-symmetric evolution between  $t_j$ and $t_k$.  We can have joint probability distribution $P(m_j, m_k)$. AOT conditions in this case demands that 
\begin{eqnarray}
\label{ptaot}
 P(m_j) \equiv \sum_{m_k = \pm 1}P(m_j, m_k)
\end{eqnarray}
Since, $t_j<t_k$, here $ P(m_j)$ is the probability to obtain the outcome $m_j$ at $t = t_j$, which can be written as 
\begin{eqnarray}
P(m_j) = Tr[\rho_j  \Pi_j ^{m_j}]
\end{eqnarray}
where $  \Pi_j ^{m_j} =  \frac{1 + m_j M_j}{2}$. The joint probability in right-hand side of Eq.(\ref{ptaot}) can be re-written as 
\begin{eqnarray}
\nonumber
\label{ptaot2}
 &&\sum_{m_k = \pm 1}P(m_j, m_k)= P(m_j, m_k = +) + P(m_j, m_k = -) \nonumber\\
 &&= Tr[U_{\alpha}(t_k-t_j) \Pi_j ^{m_j} \rho_j \Pi_j ^{m_j} U^{\dagger}_{\alpha}(t_k-t_j) \Pi_k ^{+}] +Tr[U_{\alpha}(t_k-t_j) \Pi_j ^{m_j} \rho_j \Pi_j ^{m_j} U^{\dagger}_{\alpha}(t_k-t_j)\Pi_k ^{-} ]  \nonumber\\
 && = Tr[ \Pi_j ^{m_j} \rho_j \Pi_j ^{m_j} \big(U^{\dagger}_{\alpha}(t_k-t_j)\Pi_k ^{+} U_{\alpha}(t_k-t_j)+U^{\dagger}_{\alpha}(t_k-t_j)\Pi_k ^{-} U_{\alpha}(t_k-t_j)\big)]
\end{eqnarray}
where $\Pi_k ^{\pm} = \frac{1 \pm \sigma_k}{2}$, then
\begin{eqnarray}
\label{evo}
&&U^{\dagger}_{\alpha}(t_k-t_j)\Pi_k ^{+} U_{\alpha}(t_k-t_j)+U^{\dagger}_{\alpha}(t_k-t_j) \Pi_k ^{-} U_{\alpha}(t_k-t_j) \\ \nonumber &&= U^{\dagger}_{\alpha}(t_k-t_j)(\frac{1+\sigma_k}{2}) U_{\alpha}(t_k-t_j) +U^{\dagger}_{\alpha}(t_k-t_j) (\frac{1-\sigma_k}{2}) U_{\alpha}(t_k-t_j) = U^{\dagger}_{\alpha}(t_k-t_j) U_{\alpha}(t_k-t_j)
\end{eqnarray}
By substituting Eq.(\ref{evo}) in Eq.(\ref{ptaot2}), we obtain,
\begin{eqnarray}
\label{ptaot3}
\nonumber
 \sum_{m_k = \pm 1}P(m_j, m_k) = Tr[ \Pi_j ^{m_j} \rho_j \Pi_j ^{m_j} U^{\dagger}_{\alpha}(t_k-t_j) U_{\alpha}(t_k-t_j)\\
\end{eqnarray}
Note that, if evolution is unitary implying $U^{\dagger}_{\alpha}(t_k-t_j) U_{\alpha}(t_k-t_j) = \mathbb{I}$, then for Eq.(\ref{ptaot3}), we have 
\begin{eqnarray}
\label{ptaot5}
  \sum_{m_k = \pm 1}P(m_j, m_k)  = Tr[\rho_j  \Pi_j ^{m_j}] = P(m_j) 
\end{eqnarray}
This implies that the AOT condition is satisfied in unitary dynamics. However, for $PT$ symmetric evolution, in general, $ U^{\dagger}_{\alpha}(t_k-t_j) U_{\alpha}(t_k-t_j) \neq \mathbb{I}$. As it can be seen from Eq.(\ref{evo}) that
\begin{eqnarray}
\label{utp}
U_{\alpha}(t)U^{\dagger}_{\alpha}(t) = \left(
\begin{array}{cc}
d_1 & i d_2\\
 - i d_2 & d_3 \\
\end{array}
\right)
\end{eqnarray}
where, $d_1 = \sec ^2(\alpha ) \left(\cos ^2(t-\alpha )+\sin ^2(t)\right), d_2 =2  \sec (\alpha ) \sin ^2(t) \tan (\alpha ) $ and $d_3 = \sec ^2(\alpha ) \left(\cos ^2(t+\alpha )+\sin ^2(t)\right)  $. From the matrix in Eq.(\ref{utp}), it can be seen that $U_{\alpha}(t)U^{\dagger}_{\alpha}(t) =  \mathbb{I} $ iff \(\alpha =0\) or \(t =0\). Noted that, \(\alpha =0\) leads to a unitary and \(t=0\) implies no evolution. As already mentioned, in   $PT$ symmetric evolution we have
\begin{eqnarray}
\label{ptaot4}
 P(m_j) \neq \sum_{m_k = \pm 1}P(m_j, m_k) 
\end{eqnarray}
This thereby implies that the AOT condition may be violated if the system evolves under  $\mathcal{PT}$-symmetric Hamiltonian. We demonstrate here that the quantum violation of L$\ddot{u}$der bounds of standard and variant of LGIs can be achieved when the system evolves under $PT$ symmetric Hamiltonian between two measurements in a three-time LG scenario. 

\section{Violation of LGIs using $PT$ symmetric Hamiltonian}
For our purpose, let us consider that  at $t=0$ the  state of the system is in a maximally mixed state $\rho(0) = \mathbb{I}/2$. Which evolves under $\mathcal{PT}$-symmetric non-unitary operator $U_{\alpha}(t_1)$. The state at $t = t_1$ becomes
 \begin{eqnarray} 
\rho(t_1, \alpha) = \frac{U_{\alpha}(t_1) \rho(0) U^{\dagger}_{\alpha}(t_1)}{Tr[U_{\alpha}(t_1) \rho(0) U^{\dagger}_{\alpha}(t_1)]}
\end{eqnarray}
Next, we assume the measurement of Pauli's observable $\sigma_y$ at three different times $t_1$, $t_2$ and $t_3$. The reduced state after the measurement of $t_1$ and $t_2$ will be evolved under  $PT$ symmetric evolution.
The pair-wise correlations in $L_{13}$ given in Eq.(\ref{lgi1}) are calculated as
\begin{eqnarray}
\label{cor1}
\nonumber 
\langle M_{1} M_{2}\rangle_{\alpha} &=& \frac{1}{N_2}\big[\cos (4 t)+ 2^9 \sec ^6(\alpha ) \sin ^6(t) \cos ^4(t)+ 2^7 \sec ^4(\alpha ) \sin ^4(t) \cos ^4(t) (4 \cos (2 t)-3) \\ &+&2 \sec ^2(\alpha ) \sin ^2(t) (2 \cos (4 t)-1) (2 \cos (2 t) +2 \cos (4 t)-1) \big],
\end{eqnarray}
\begin{eqnarray}
\label{cor2}
\langle \hat{M_{2}} \hat{M_{3}}\rangle_{\alpha} &=& \frac{1}{N_3}\big[ 2^7 \sec ^6(\alpha ) \sin ^6(t) (2 \cos (t)+\cos (3 t))^2 + 8 \sec ^4(\alpha ) \sin ^4(t) (6 \cos (2 t)+10 \cos (4 t)\\ \nonumber &+&10 \cos (6 t)+4 \cos (8 t)+1)+\cos (6 t)+   4 \sec ^2(\alpha ) \sin ^2(t) (\cos (2 t)+\cos (4 t) - \cos (6 t)+\cos (8 t)+\cos (10 t)+1)\big]
\end{eqnarray}
and
\begin{eqnarray}
\label{cor3}
\langle \hat{M_{1}} \hat{M_{3}}\rangle_{\alpha} &=& \frac{1}{N_4}\big[128 \sec ^6(\alpha ) \sin ^6(t) (2 \cos (t)+\cos (3 t))^2 +8 \sec ^4(\alpha ) \sin ^4(t) (4 \cos (2 t)+12 \cos (4 t) \\ \nonumber &+&10 \cos (6 t)+ 4 \cos (8 t)-1)+\cos (4 t)-8 \sec ^2(\alpha ) \sin ^4(t) (8 \cos (2 t) + 10 \cos (4 t)+6 \cos (6 t)+2 \cos (8 t)+3) \big].
\end{eqnarray}
where $N_2, N_3$ and $N_4$ are the normalization constants given by Eq.(\ref{n1}), Eq.(\ref{n2}) and Eq.(\ref{n3}) respectively in the Appendix.

Quantum value of $L_{13}$ can be calculated by using the correlations in Eq.(\ref{cor1}-(\ref{cor3}) which we denote as 
  $L^{\alpha}_{13}$. For $\alpha=0$, $L^{\alpha}_{13}$ reduces to unitary case, i.e., $L_{13} = 2 \cos(2t)-  \cos(4t)$ as given in Eq.(\ref{qlgi}). In Figure 1, we have plotted  $L^{\alpha}_{13}$ for four different values of $\alpha$. It can be seen that for $\alpha = 0 $ the optimal value of  $L^{\alpha = 0}_{13}$  is $1.5$ which is the L$\ddot{u}$ders bound obtained in unitary QM. However, in $\mathcal{PT}$-symmetric evolution as $\alpha$ increases the violation of $L^{\alpha}_{13}$ approaches to algebraic maxima $3$ for $\alpha \approx \pi/2$. 
\begin{figure}[ht]
\includegraphics[width=0.9\linewidth]{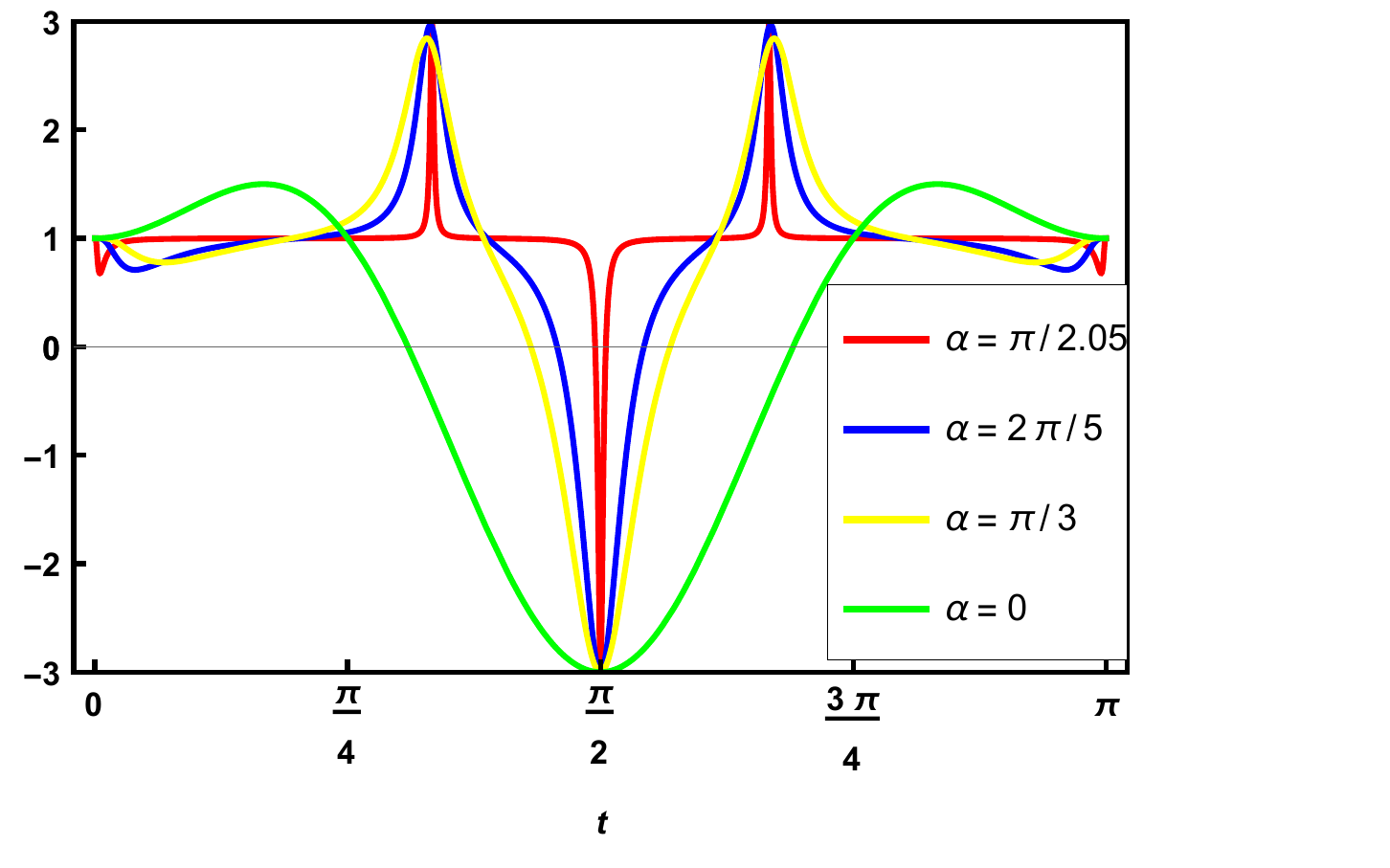}
\caption{(color online):$L^{\alpha}_{13}$ is plotted with respect to $t$ for different values  for different values $\alpha =0$,  $\alpha =\pi/3$,  $\alpha =2 \pi/5$ and $\alpha = \pi/2.05$. }
\label{fig:1}
\end{figure}
The quantum expression of variant of LGI $V_3^{\alpha}$ is very large  due to it's state dependence behavior. Therefore we skip the general expression of variant of LGI. However, we numerically found that for the state given in  Eq.(\ref{state}), the quantum value of  $V^{\alpha}_{3}$ approaches $3$ at $\theta = 5 \pi/6$, $\phi = \pi/2$ , $\alpha =\pi/3$ and $ t = 0.785$. In Figure 2, we have plotted the variant of LGI  $V^{\alpha}_{3}$ for four different values of $\alpha$. It can be seen that for $\alpha = 0 $ the optimal value is $1.93$ and as $\alpha$ increases the violation of $V^{\alpha}_{3}$ approaches to algebraic maxima $3$ for $\alpha \approx \pi/2$.\\
\begin{figure}[ht]
	\includegraphics[width=0.9\linewidth]{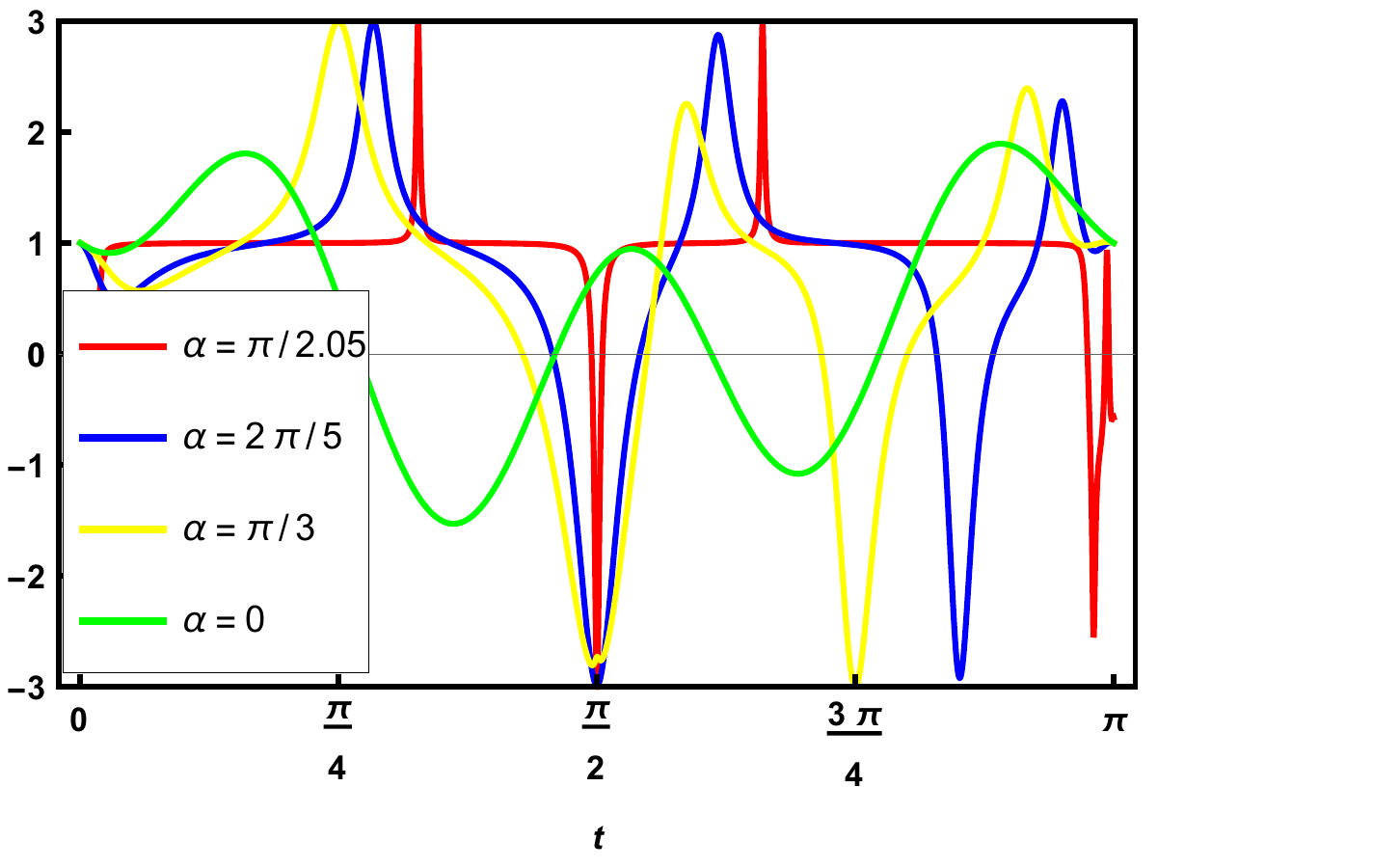}
	\caption{(color online):$V^{\alpha}_{3}$ is plotted with respect to $t$ for different values $\alpha =0$,  $\alpha =\pi/3$,  $\alpha =2 \pi/5$ and $\alpha = \pi/2.05$ . }
	\label{fig:1}
\end{figure}
In order to understand this feature, we provide a detailed analysis of the assumption of LGI in the next section.

\section{Quantification of no-signaling-in-time and Arrow-of-time}
To critically examine the implications for the quantum violations of L$\ddot{u}$ders bound of standard and variant of LGIs using $\mathcal{PT}$-symmetric evolution, we define the degree of NSIT and AOT violations. The NSIT (AOT) condition assumes that the probability of obtaining an outcome of the measurements remains unaffected due to the prior (posterior) measurements. The conjunction of all the NSIT and AOT conditions ensures the existence of the global joint probability distribution $P(m_1, m_2,  m_3)$. The two-time NSIT conditions are given by
\begin{eqnarray}
NSIT_{(i)j} : P(m_j) = \sum_{m_1=\pm 1}P({m_i}, m_j)
\end{eqnarray}
where, $i<j$. Similarly three-time NSIT conditions are given by
\begin{eqnarray}
NSIT_{1(2)3}:P(m_1, m_3)  \equiv \sum_{m_2 = \pm 1}P(m_1, m_2, m_3) 
\end{eqnarray}  
and
\begin{eqnarray}
NSIT_{(1)23}:P(m_2, m_3) \equiv \sum_{m_1 = \pm 1}P(m_1, m_2, m_3)
\end{eqnarray}
The AOT conditions can be written as
\begin{eqnarray}
AOT_{i(j)}: P(m_i) \equiv \sum_{m_j = \pm 1}P(m_i, m_j)
\end{eqnarray}
\begin{eqnarray}
AOT_{12(3)}: P(m_1, m_2)  \equiv \sum_{m_3 = \pm 1}P(m_1, m_2, m_3)
\end{eqnarray}
As already mentioned that a suitable set of NSIT and AOT conditions provides the necessary and sufficient condition for macrorealism \cite{clemente15}. Then 
\begin{eqnarray}
\nonumber
NSIT_{(1)2} \wedge NSIT_{1(2)3} \wedge NSIT_{(1)23} \wedge AOT_{12(3)} \Leftrightarrow MR
\end{eqnarray}
However, NSIT conditions are necessary for LGIs, and the violation of an LGI requires at least one of the NSIT condition to be violated, but the violation of one or all NSIT conditions do not warrant the violation of a given LGI. We can then write
\begin{eqnarray}
\nonumber
NSIT_{1(2)3} \wedge NSIT_{(1)23} \wedge AOT_{12(3)} \Rightarrow LGIs
\end{eqnarray}
The NSIT conditions can be shown to be violated within unitary QM but AOT conditions are always satisfied. We first show here that how LG expressions can be cast in terms of degrees of violations of NSIT and AOT conditions. For this let us define the degree of violations of various NSIT and AOT conditions, So, degree of violation for $NSIT_{(1)23}$ and $NSIT_{1(2)3}$ are quantified as
\begin{eqnarray}
\nonumber
\label{d11}
&&D_{(1)23}(m_2,m_3) = P({m_2},{m_3})-\sum_{m_1 = \pm 1} P({m_1},{m_2}, {m_3})\\
 \\
\label{d22}
\nonumber
&&D_{1(2)3}(m_1 ,m_3) = P(m_1, m_3)-\sum_{m_2 = \pm 1} P(m_1, m_2, m_3)\\
\end{eqnarray}
respectively. $D_{(1)23}(m_2,m_3)$ ($D_{1(2)3}(m_1,m_3)$) is the amount of disturbance (degree of violation of NSIT conditions) created by the measurement $M_1$($M_2$) at $t_1$ ($t_2$) to the measurements of $M_2$ and $M_3$ ($M_1$ and $M_3$) at $t_2$ and $t_3$ ($t_1$ and $t_3$) respectively. 

The degree of violation of $AOT_{12(3)}$ and$AOT_{1(23)}$  are quantified as
\begin{eqnarray}
\nonumber
\label{d33}
R_{12(3)}(m_1,m_2)&=&P(m_1,m_2)-\sum_{m_3 = \pm 1} P(m_1, m_2, m_3)\\
\end{eqnarray}
\begin{eqnarray}
\nonumber
\label{d331}
R_{1(23)}(m_1)&=&P(m_1)-\sum_{m_2, m_3 = \pm 1} P(m_1, m_2, m_3)\\
\end{eqnarray}
respectively.

 Note that $ D_{(1)23}(m_2, m_3)$,  $D_{1(2)3}(m_1, m_3)$, $R_{12(3)}(m_1, m_2)$ and $R_{1(23)}(m_1)$ can take both positive and negative values. Henceforth, for avoiding the clumsiness of the notation in disturbance inequalities of LGIs, we denote $ D_{(1)23}(m_2, m_3),  D_{1(2)3}(m_1, m_3)$, $R_{12(3)}(m_1,m_2)$ and $R_{1(23)}(m_1) $ by $D_{(1)23}$, $D_{1(2)3}$, $R_{12(3)}$ and $R_{1(23)} $ respectively.
\begin{widetext}
\begin{center}
\begin{figure}[ht]
\includegraphics[width=0.9\linewidth]{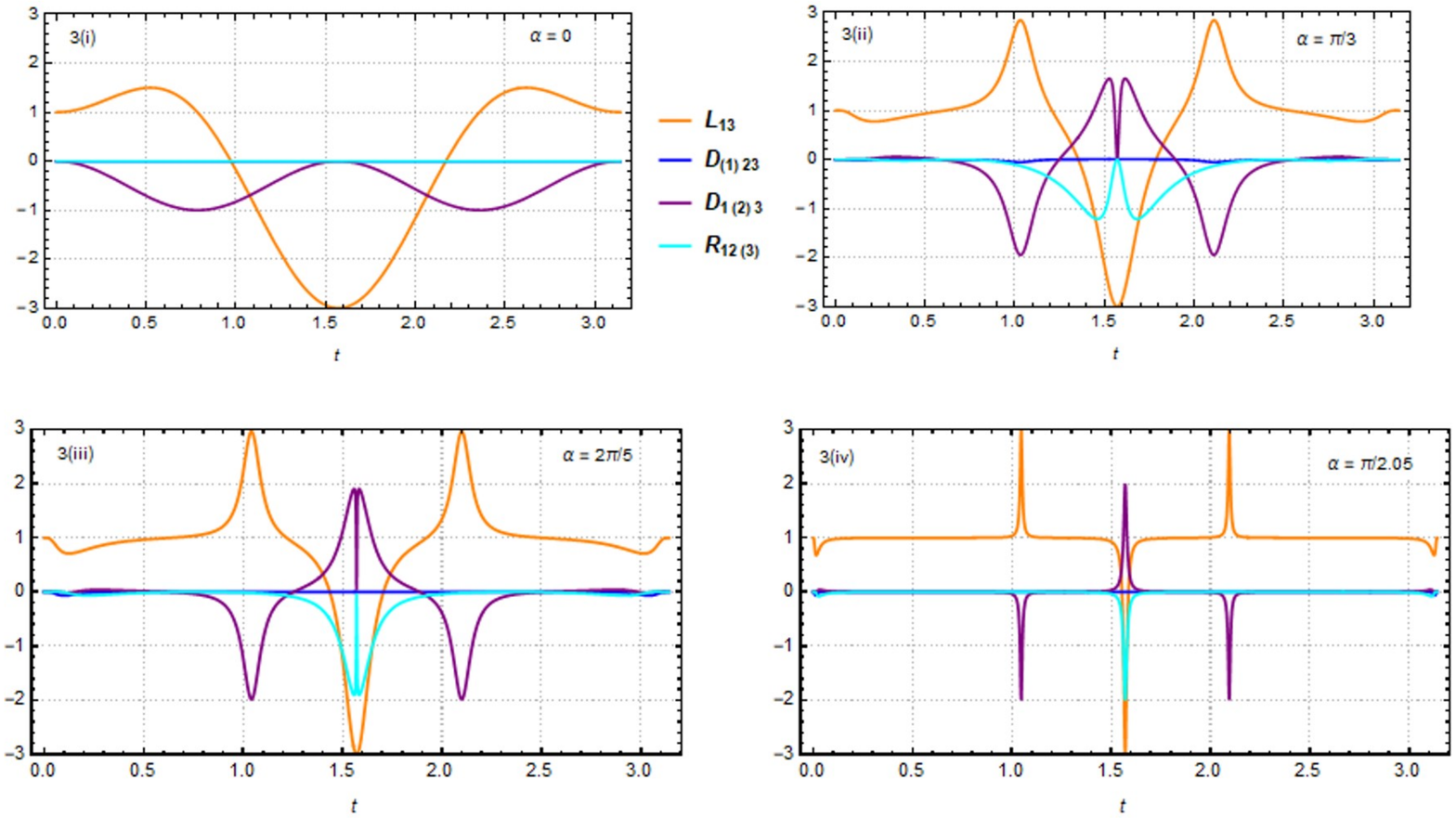}
\caption{(color online): The quantum expression of $L_13,\alpha$,, $D_{(1)23}$, $D_{1(2)3}$ and $R_{12(3)}$  are plotted with respect to $t$ for $\alpha = 0$, $\alpha = \pi/3$, $\alpha = 2 \pi/5$ and $\alpha = \pi/2.05$ in Fig.3(i), 3(ii), 3(iii) and 3(iv) respectively.}
\label{fig:1}
\end{figure}
\end{center}

\begin{center}
		\begin{figure}[ht]
			\includegraphics[width=0.9\linewidth]{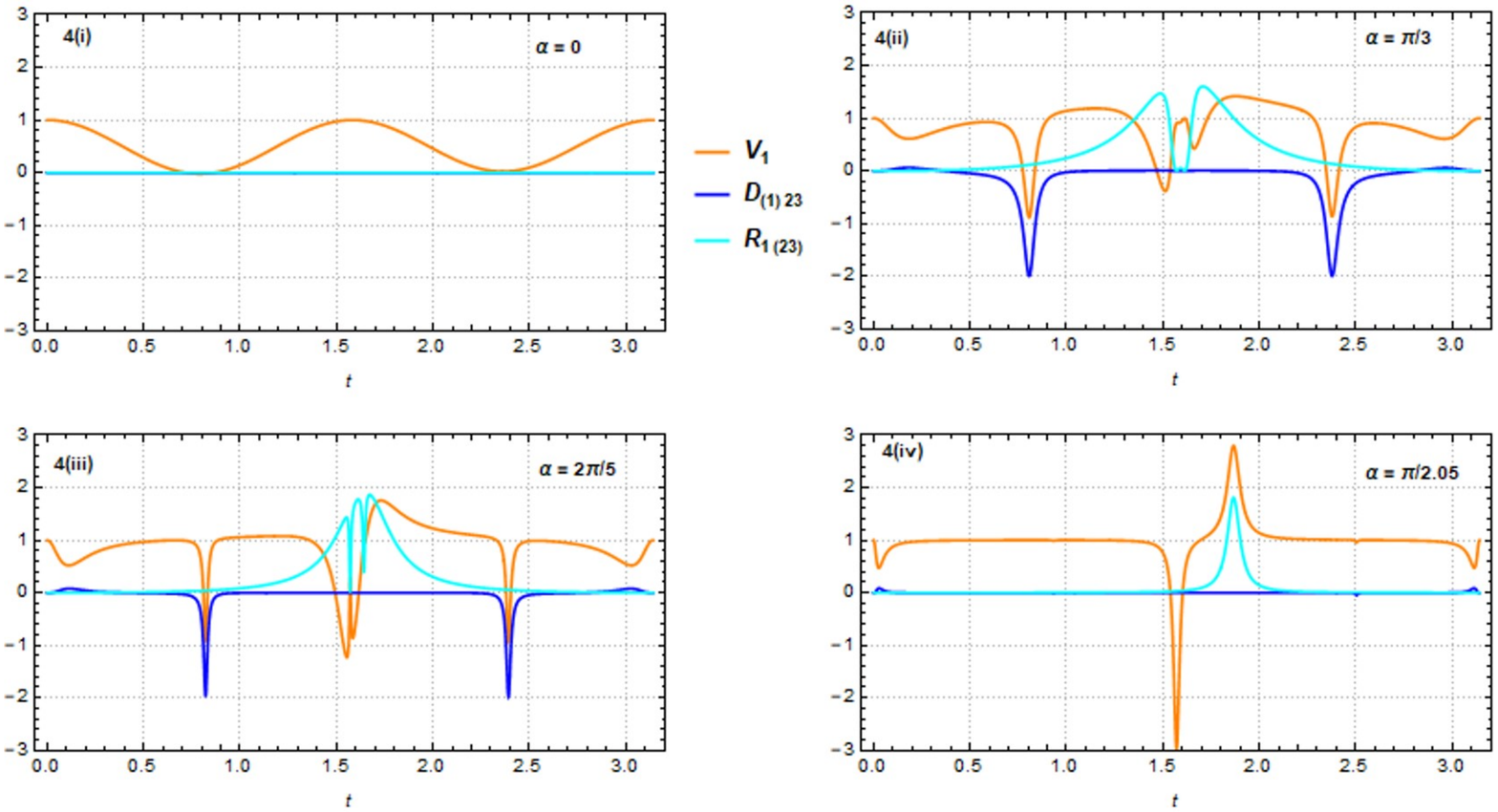}
			\caption{(color online): The quantum expression of $V^{\alpha}_1$, $D_{(1)23}$ and $R_{1(23)}$  are plotted with respect to $t$ for $\alpha = 0$, $\alpha = \pi/3$, $\alpha = 2\pi/5$ and $\alpha = \pi/2.05$ in Fig.4(i), 4(ii), 4(iii) and 4(iv) respectively.}
			\label{fig:1}
		\end{figure}
	\end{center}
\end{widetext}
Appendix B discusses the standard and variant of LG expressions in terms of the degree of NSIT and AOT conditions violations.

The quantum expression of $L^{\alpha}_{13}$ (obtained impinging in Eq.(\ref{lgi1})), $D_{(1)23}$, $D_{1(2)3}$ and $R_{12(3)}$  with respect to $t$ for different value of $\alpha$ are plotted in Figure 3. From Figure 3 it can be seen that along with the violation of at least one of the NSIT conditions there is violation of AOT condition $R_{12(3)}$ too.

 In order to understand the importance of AOT condition in LG scenario evolved under $\mathcal{PT}$-symmetric evolution, we further study the degree of violation of NSITs and AOT conditions for the case of a variant of LGI given by
\begin{eqnarray}
\label{v1g}
V_{1} &=&  -\langle M_{1} M_{2} M_{3}\rangle + \langle M_{2} M_{3}\rangle +\langle M_{1}\rangle \leq 1
\end{eqnarray}

The quantum expression of $V^{\alpha}_1$,  $D_{(1)23}$ and $R_{1(23)}$ is quite lengthy so that we skip including that here.  The quantities $V^{\alpha}_1$,  $D_{(1)23}$ and $R_{1(23)}$  are plotted with respect to $t$ for $\alpha = 0$, $\alpha = \pi/3$, $\alpha = 2\pi/5$ and $\alpha = \pi/2.05$ in Figure 4. It can be seen that as $\alpha$ increase the range of $t$ showing the violation of $V^{\alpha}_1$ with \textit{no} violation of NSIT condition but with the violation of AOT condition is increasing. Finally at $\alpha =\pi/2.05$ there is no violation of NSIT condition but AOT condition is violated.

The above findings show that evolution governed by $\mathcal{PT}$-symmetric Hamiltonian violates the arrow-of-time condition. The reason for the violation of this fundamental principle of QM can be understood by considering the post-measurement density matrix of the system. Note that the post-measurement density matrix in $\mathcal{PT}$-symmetric QM is not normalized as in the case of standard QM. Resultant, in order to find one-time, two-time, or three-time correlation functions in case of standard and variant of LGIs with the system state evolved under $\mathcal{PT}$-symmetric evolution, we have to re-normalize the reduced density matrix of the system again and again. This indicates the absence of the second assumption on which  $\mathcal{PT}$ -symmetric  quantum is built \cite{b01}.

\section{Summary and Discussion}
To summarize, we have studied the quantum violations of two formulations of LGIs when the system evolves under $\mathcal{PT}$-symmetric Hamiltonian in between the measurements at two different times. The $\mathcal{PT}$-symmetric systems exhibit intriguing behavior around the exceptional points  $(\alpha\approx\pi/2$), the points of coalescence of both eigenvalues as well as eigenvectors. This paper considers the standard LGI and a variant of LGI in a three-time LG scenario. The respective L$\ddot{u}$der bounds of standard and variant of LGIs are 1.5 and 2. We first demonstrated that the quantum violation of both the standard and variant of LGIs approach algebraic maximum near the exceptional points, thereby violating the L$\ddot{u}$der bounds.  

To critically examine the implications for the quantum violations of L$\ddot{u}$ders bound of standard and variant of LGIs for the case of $\mathcal{PT}$-symmetric evolution, we introduced the degree of NSIT and AOT violations. Unlike the unitary dynamics, in $\mathcal{PT}$-symmetric evolution, there is a possibility of violating the AOT condition near the exceptional points.   It is already known \cite{clemente15,swati17} that for unitary dynamics, the quantum violation of a LGI requires one of the NSIT conditions to be violated, and the AOT condition is naturally satisfied. Here we cast the LGIs in terms of the degree of NSIT and AOT conditions which enables us to show that the violation of  L$\ddot{u}$ders bound is obtained when both NSIT and AOT conditions are violated. Interestingly, for the case of the variant of LGIs, for a suitably chosen range of relevant parameters all the NSIT conditions are satisfied and the violation of L$\ddot{u}$ders bound can solely be explained from the violation of AOT condition alone. This feature is explicitly shown in Fig. 4. Hence, the quantum violation of variant of LGI implies a disturbance caused by future measurement to the past measurements.

\begin{widetext}
\appendix
\section{}
For the initial system state $\rho_0 = I/2$ evolve under $\mathcal{PT}$-symmetric non unitary dynamics $U_{\alpha}(t_1)$ normalization constants of pair-wise correlations $\langle M_{1} M_{2}\rangle_{\alpha} $, $\langle M_{2} M_{3}\rangle_{\alpha}$ and $\langle M_{1} M_{3}\rangle_{\alpha} $  of $L_{13}$ in Eq.(\ref{lgi1}) are given by
\begin{eqnarray}
\label{n1}
N_{1} &=& \frac{1}{2} \bigg(1024 \sec ^6(\alpha ) \sin ^6(t) \cos ^4(t)+64 \sec ^4(\alpha ) \sin ^4(t) \cos ^2(t) (2 \cos (2 t)+5 \cos (4 t)-1)\nonumber\\ &+&4 \sec ^2(\alpha ) \sin ^2(t) (2 \cos (2 t)-\cos (4 t)+4 (\cos (6 t)+\cos (8 t)+1))+\cos (2 t)+\cos (10 t)\bigg),
\end{eqnarray}
\begin{eqnarray}
\label{n2} 
N_{2} &=& 128 \sec ^6(\alpha ) \sin ^6(t) (2 \cos (t)+\cos (3 t))^2+8 \sec ^4(\alpha ) \sin ^4(t) \cos (4 t) (24 \cos (2 t)+10 \cos (4 t)+15) \nonumber\\ &+&2 \sec ^2(\alpha ) \sin ^2(t) (2 \cos (4 t)-1) (7 \cos (2 t)+4 \cos (4 t)+4 \cos (6 t)+3)+\cos ^2(6 t)
\end{eqnarray}
and
\begin{eqnarray}
\label{n3}
N_{3} &=& \frac{1}{2} \bigg(256 \sec ^6(\alpha ) \sin ^6(t) (2 \cos (t)+\cos (3 t))^2+16 \sec ^4(\alpha ) \sin ^4(t) (6 \cos (2 t)+13 \cos (4 t)+12 \cos (6 t)\nonumber\\&+&5 \cos (8 t)+1)+4 \sec ^2(\alpha ) \sin ^2(t) (7 \cos (2 t)+\cos (6 t)+4 (\cos (8 t)+\cos (10 t)+1))+\cos (4 t)+\cos (12 t)\bigg)
\end{eqnarray}
respectively.

\section{}
In a macrorealist theory, the LGIs are derived by considering $D_{(1)23}(m_2, m_3)= D_{1(2)3}(m_1, m_3) = 0$ and $R_{12(3)}(m_1, m_2) = 0$ is assumed to be naturally satisfied. In QM it can be seen that the violation of LGI is achieved if either or both of $ D_{(1)23}(m_2, m_3)$ and $D_{1(2)3}(m_1, m_3)$ are non-zero (violation of NSIT condition). It can be understood through the study of the difference of $L_{13}$ and $L_{123}$, where the expression of $L_{123}$, obtained by considering all the three measurements for each correlations in $L_{13}$ given in Eq.(\ref{lgi1}), can be written as
\begin{eqnarray}
\label{lg123}
L_{123} &=&  \langle M_{1} M_{2}\rangle_{123} + \langle M_{2} M_{3}\rangle_{123} -\langle M_{1} M_{3}\rangle_{123} =  1 - 4 \beta
\end{eqnarray}
where, $\beta = P(M_{1}^{+} M_{2}^{-} M_{3}^{+})+P(M_{1}^{-} M_{2}^{+} M_{3}^{-})$,
\begin{eqnarray}
 \langle M_{1} M_{2}\rangle_{123} = \sum_{m_1, m_2 = \pm 1} m_1 m_2 \sum_{m_3 = \pm 1} m_3 P(m_1, m_2, m_3) 
\end{eqnarray}
and similarly for $\langle M_{2} M_{3}\rangle_{123}$ and $\langle M_{1} M_{3}\rangle_{123}$. 

Now, using Eq.(\ref{d11}-\ref{d33}), we can write 
\begin{eqnarray}
L_{13} - L_{123}&=&\sum_{m_2= m_3}D_{(1)23}(m_2, m_3)  -\sum_{m_2 \neq m_3}D_{(1)23}(m_2, m_3)  +\sum_{m_1 = m_2}R_{12(3)}(m_1, m_2)-\sum_{m_1 \neq m_2}R_{12(3)}(m_1, m_2)\\ \nonumber&&+\sum_{m_1 \neq m_3}D_{1(2)3}(m_1, m_3)-\sum_{m_1 = m_3}D_{1(2)3}(m_1, m_3)\\
\end{eqnarray}
Since $L_{13} \leq 1$ and $L_{123} = 1- 4 \beta $ we obtain
\begin{eqnarray}
\nonumber
\label{NLgi1}
\sum_{m_2 = m_3}D_{(1)23}(m_2, m_3)-\sum_{m_1 = m_3}D_{1(2)3}(m_1, m_3)+\sum_{m_1 = m_2}R_{12(3)}(m_1, m_2) \leq 2 \beta
\end{eqnarray}
We have thus written the standard LG expression in terms of the degrees of NSIT and AOT violations. Hence, for the violation of LGI, one needs
\begin{eqnarray}
\nonumber
\label{Lgiv1}
\sum_{m_2 = m_3}D_{(1)23}(m_2, m_3)-\sum_{m_1 = m_3}D_{1(2)3}(m_1, m_3) +\sum_{m_1 = m_2}R_{12(3)}(m_1, m_2) > 2 \beta
\end{eqnarray}
should be satisfied. Hence, we can say that the violation of standard LGI depends on interplay between the violations of NSITs, AOT conditions and a threshold value  $2 \beta$. Next, we consider the variant of LGI in Eq.(\ref{vlgi1}). In order to cast the variant of LGI in terms of degree of violation of NSIT and AOT conditions by following the step earlier we consider the difference between $V_1$ and $V_{123}$.
  The expression of $V_{123}$, obtained by considering the measurements of all the three observables for each correlations in $V$ given in Eq.(\ref{vlgi1}). Thus $V_{123}$ can be written as
\begin{eqnarray}
\label{vlg123}
V_{123} = - \langle M_{1} M_{2} M_{3}\rangle_{123} + \langle M_{2} M_{3}\rangle_{123} +\langle M_{1}\rangle_{123}  =  1 - 4 \delta
\end{eqnarray}
where, $\delta = P(M_{1}^{-} M_{2}^{+} M_{3}^{-})+P(M_{1}^{-} M_{2}^{-} M_{3}^{+})$ and
\begin{eqnarray}
\langle M_{1} M_{2} M_{3}\rangle_{123} = \sum_{m_1,m_2, m_3 = \pm 1} m_1 m_2 m_3 P(m_1, m_2, m_3),  \langle M_{1}\rangle_{123} =  \sum_{m_1  = \pm 1} m_1 \sum_{m_2, m_3 = \pm 1} m_2 m_3 P(m_1, m_2, m_3)
\end{eqnarray}
Now $\langle M_{1} M_{2} M_{3}\rangle_{123} = \langle M_{1} M_{2} M_{3}\rangle$. The NSIT condition for $V_1$ is given in Eq.(\ref{d11}). AOT condition for $V_1$ is quantified as
\begin{eqnarray}
\label{Im2}
R_{1(23)}(m_1)=P(m_1)-\sum_{m_2,m_3 = \pm 1} P(m_1,m_2, m_3)
\end{eqnarray}
respectively. Using Eq.(\ref{d11}) and Eq.(\ref{Im2}) we can write
\begin{eqnarray}
\label{vv123}
V_1 - V_{123} &=& 2\sum_{m_2= m_3}D_{(1)23}(m_2, m_3)-R_{1(23)}(m_1)
\end{eqnarray}
By noting $V \leq 1$ and substituting $V_{123} = 1 - 4 \delta$ from Eq.(\ref{vlg123}), we obtain
\begin{eqnarray}
2\sum_{m_2= m_3}D_{(1)23}(m_2, m_3)-R_{1(23)}(m_1) \leq 4 \delta
\end{eqnarray}
Then for the violation of variant of LGI $V_1$,
\begin{eqnarray}
\label{v1}
2\sum_{m_2= m_3}D_{(1)23}(m_2, m_3)-R_{1(23)}(m_1)  \geq 4 \delta
\end{eqnarray}
From Eq.(\ref{v1})  we can say that the violation of variant of LGI depends on the interplay between degree of violation of NSIT and AOT conditions.

\end{widetext}
\end{document}